\documentclass[structabstract]{aa}  

\usepackage{graphicx}

\usepackage{txfonts}

\begin{document}

\title{Linear and nonlinear MHD mode coupling of the fast magnetoacoustic wave about a 3D magnetic null point}


   \author{J.~O. Thurgood
          \and
          J.~A. McLaughlin}

   \institute{School of Computing, Engineering \& Information Sciences, Northumbria University, Newcastle Upon Tyne, NE1 8ST, UK}

\offprints{J.~O. Thurgood, \email{jonathan.thurgood@northumbria.ac.uk}}


\date{Received $20$ June $2012$ / Accepted $29$ July $2012$}

 
  \abstract
   {Coronal magnetic null points have been implicated as possible locations for localised heating events in 2D models. We investigate this possibility about fully 3D null points.}
   {We investigate the nature of the fast magnetoacoustic wave about a fully 3D magnetic null point, with a specific interest in its propagation, and we look for evidence of MHD mode coupling and/or conversion to the Alfv\'en mode.}
   {A special fieldline and flux-based coordinate system is constructed to permit the introduction of a pure fast magnetoacoustic wave in the vicinity of  proper  and improper 3D null points. We consider the ideal, $\beta=0$,  MHD equations which are solved using the {\emph{LARE3D}} numerical code. The constituent modes of the resulting wave are isolated and identified using the special coordinate system. Numerical results are supported by analytical work derived from perturbation theory and a linear implementation of the WKB method.}
   {\emph{An initially pure fast wave is found to be permanently decoupled from the Alfv\'en mode both linearly and nonlinearly  for both proper and improper 3D null points}. The pure fast mode also generates and sustains a nonlinear disturbance aligned along the equilibrium magnetic field. The resulting pure fast magnetoacoustic pulse has transient behaviour  which is found to be governed by the (equilibrium) Alfv\'en-speed profile, and a refraction effect focuses all the wave energy towards the null point. }
   {Thus, the main results from previous 2D work do indeed carry over to the fully 3D magnetic null points and so we conclude that {\emph{3D null points are    locations for preferential heating in the corona by 3D fast magnetoacoustic waves}}.}


\keywords{Magnetohydrodynamics (MHD) -- Waves -- magnetic fields -- Sun:~corona -- Sun:~ magnetic topology -- Sun:~oscillations  }

\titlerunning{Linear and nonlinear MHD mode coupling at a 3D null point.}

\authorrunning{Thurgood \& McLaughlin}
   \maketitle


\newpage

\section{Introduction}\label{section:1}
An abundance of observational data from solar instruments including {\emph{SOHO}} (e.g. Ofman et al. \cite{ofman.etal.1997}; DeForest \& Gurman \cite{plumes}), {\emph{TRACE}} (e.g.  Nakariakov {{et al.}} \cite{Nakariakov1999}; De Moortel et al. \cite{demoortel.etal.2000}) and {\emph{Hinode}} (e.g. Ofman \& Wang \cite{ofman.wang.2008}) confirms the existence of magnetohydrodynamic (MHD) wave motions in the coronal plasma (see also reviews by De Moortel \cite{ineke2005};  Nakariakov \& Verwichte \cite{Nakariakov2005}; Ruderman \& Erd\'elyi \cite{Misha2009};    Goossens et al. \cite{goosens2011}; De Moortel \& Nakariakov \cite{ineke2012}). MHD wave theory suggests that the corona potentially supports a wide variety of distinct classes of wave motions, namely the Alfv\'en wave, and fast and slow magnetoacoustic waves. However, the applicability and appropriateness of such classifications has recently provoked intense discussion, for instance reports of having  observed Alfv\'en waves in the corona (Tomcyzyk et al. \cite{tomczyk.etal.2007}; De Pontieu et al. \cite{Bart2007}) are contested by, e.g., Erd\'elyi \& Fedun (\cite{Erdelyi2007}) and Van Doorsselaere et al. (\cite{Tom2008}).

The \lq{classic}\rq{ } terminology of MHD waves originates in the analysis of modes supported by magnetically-unidirectional, homogeneous plasmas of infinite extent which consider either plane wave solutions or utilise the method of characteristics (viz. Riemann decomposition) (examples of such analyses can be found in, e.g. Friedrichs \& Kranzer \cite{friedrichs}; Lighthill \cite{lighthill}; Cowling \cite{Cowling1976}; Goedbloed \& Poedts \cite{goedbloedbook1}). Here, three distinct modes are permitted, the Alfv\'en mode, and the fast and slow magnetoacoustic, and their behaviour and nature is well understood. For a low-$\beta$ plasma, it is found that the Alfv\'en wave is a transverse, purely-magnetic wave, propagating at the Alfv\'en speed, guided by the magnetic field. The fast magnetoacoustic wave is found to propagate roughly isotropically  at the fast speed ($c_F=\sqrt{c^{2}_{A}+c^{2}_S}$, $c_F\approx c_A$ where $\beta\ll1$), and can travel along and across magnetic fieldlines. The slow magnetoacoustic wave propagates longitudinally along the magnetic fieldlines, roughly at the sound speed.

However in the solar corona,  effects including gravitational stratification, inhomogeneous density profiles and multiple-source magnetic field geometries call into question whether these three, classical modes are still valid. The departure to inhomogeneity typically introduces a variety of phenomena such as resonant absorption (in the corona see, e.g., Ionson \cite{Ionson78}; \cite{Ionson82};  \cite{Ionson83}; Hollweg \cite{Hollweg1984}; Ruderman \& Roberts \cite{rudermanroberts}) and phase mixing (e.g. Heyvaerts \& Priest \cite{HP1983}), all of which blur the distinction between these separate modes (see also, e.g.,  Bogdan et al. \cite{bogdan}; McDougall \& Hood  \cite{Dee2007}; Sousa \& Cunha \cite{Sousa2008}; Cally \& Hansen \cite{Cally2011}; Hansen \& Cally \cite{Hansen2012}).

Typically theorists preserve the concept of the three MHD modes and introduce the concept of conversion and coupling between the constituent modes as waves encounter certain inhomogeneous features. However, as highlighted by Goossens et al. (\cite{goosens2011}), we must take care to understand that realistic plasmas, strictly speaking, do not support three \lq{classic}\rq{ } wave modes but rather that a propagating MHD pulse encounters situations in which it assumes transient properties that qualitatively correspond to one (or more) of the homogeneous modes. 
Conversely, it should also be possible to locally determine hyperbolic characteristics corresponding to the fast, slow and Alfv\'en waves for any ideal MHD system via Riemann decomposition (for an overview see, e.g., Goedbloed et al. \cite{goedbloedbook2})
although, as discussed in Goedbloed \& Poedts (\cite{goedbloedbook1}), this is not always analytically possible for fully inhomogeneous cases (in particular for non-unidirectional magnetic fields).

As such, it is unclear whether or not the classic MHD modes exist in the solar atmosphere as discrete entities or if such mode separation requires specific geometries. For example, as discussed by Parker (\cite{parker91}), the existence of true Alfv\'en waves as per Alfv\'en (\cite{alfven}) is dependent upon special magnetic geometries which contain invariant directions. Despite this uncertainty,
a theoretical framework which involves them conceptually is not necessarily obsolete. 	Mode interpretation and analysis that relates  MHD wave behaviour in inhomogeneous, realistic plasmas to the comparatively simple, classic waves still has the potential to be a useful framework.

In this paper, we are specifically interested with the behaviour of MHD waves in the vicinity of the 3D {\emph{magnetic null point}} (e.g. Parnell et al. \cite{Parnell1996}; Priest \& Forbes \cite{magneticreconnection2000}). Null points are topological  features of coronal magnetic fields, predicted by magnetic field extrapolations such as Beveridge et al. (\cite{BPB}) and Brown \& Priest (\cite{BP}). At these magnetic null points, the magnetic induction is zero. Hence, approaching a magnetic null point, the coronal plasma is highly inhomogenous.
In the solar atmosphere, null points have been identified as playing key roles in many processes, for example; in  CMEs (in the \emph{magnetic breakout model}, e.g. Antiochos \cite{Antio98}; Antiochos et al. \cite{Antio99}), magnetic reconnection (e.g. Priest \& Forbes \cite{magneticreconnection2000}) and oscillatory reconnection (reconnection driven by wave-null interactions, see McLaughlin et al. \cite{MDHB2009}; McLaughlin et al. \cite{james2012}) all of which are thought to play a role in coronal heating.
As both waves and null points are ubiquitous in the corona (Close et al. \cite{Close04}, Longcope \& Parnell \cite{Longcope09} and R\'egnier et al. \cite{Regnier08} give a rough estimates of $1.0-4.0\times10^{4}$ null points) wave-null interactions are inevitable and thus are arguably a fundamental plasma process in the solar atmosphere.
Hence, our present research is specifically concerned with the extension of classic MHD wave theory  about null points,  which in a broader sense contributes to both the theory of MHD wave behaviour in inhomogeneous media and the understanding of fundamental aspects of coronal physics.

MHD wave behaviour in the neighbourhood of magnetic null points have been extensively studied in 2D models. Bulanov \& Syrovatskii (\cite{Bulanov1980}) performed the first investigation of MHD behaviour about a 2D null and noted that in a 2D geometry the  motions governing the Alfv\'en mode and the fast magnetoacoustic modes are decoupled, permitting analysis that considers the modes separately.  The transient features of fast and Alfv\'en waves in various 2D null point geometries within a $\beta=0$ plasma were studied extensively in a series of papers by McLaughlin \& Hood (\cite{MH2004}; \cite{MH2005}; \cite{MH2006a}). Again, these authors found the two modes decoupled and identified key propagation features for each mode. The fast wave propagates isotropically (at the same characteristic speed both across and along magnetic field lines), with behaviour dictated by the Alfv\'en-speed profile, propagating from regions of high to low Alfv\'en speed resulting in a refraction effect which focuses the wave energy into the null point.
Meanwhile, the Alfv\'en wave is confined to follow magnetic fieldlines, thus leading the wave energy to accumulate along the separatrices.
Due to the resulting current build-up in these regions, these papers concluded that magnetic null points are likely locations for localised heating events in the corona.
 $\beta \neq 0$ and nonlinear behaviour has also been investigated by McLaughlin \& Hood (\cite{MH2006b}) and McLaughlin et al. (\cite{MDHB2009}) respectively. A comprehensive overview of the whole topic is given in the review paper of McLaughlin et al. (\cite{james2011a}).

Of course, 2D models only give an initial grounding in the physics of realistic null points and for a full understanding we must turn to 3D models  (as singularities, null points are inescapably 3D and any 2D X-point configuration in fact only captures the physics of a {\emph{null line}} of infinite extent). It is not clear to what extent the characteristics and behaviour of waves about 2D null points transfer to the fully 3D  case, and surprisingly few papers have been written that address MHD wave behaviour about a fully 3D null point. Most papers have focused on the dynamics of current accumulation over time (in an attempt to locate regions where reconnection is most likely to occur) rather than  focus on the transient propagation features of the individual modes.

Galsgaard et al. (\cite{Galsgaard2003}) consider a proper, $\beta=0$ null point and introduce a twist wave (what they call an Alfv\'en wave) which is generated about the spine and eventually accumulates on the {\emph{fan}} (perhaps analogous to Alfv\'en waves behaviour around 2D null points). In addition, they observe a small amount of current accumulation at the null itself, which they suggest is due to nonlinear generation of a fast wave. No linear coupling between wave modes is observed, and in their linear analysis they find that the wave equations for the fast and Alfv\'en modes decouple. However, this is not surprising since their proper 3D null has azimuthal symmetry, and so the system is actually 2.5D, not fully 3D.

Pontin \& Galsgaard (\cite{PG2007}), Pontin et al. (\cite{PBG2007}) and Galsgaard \& Pontin (\cite{klaus2011a}; \cite{klaus2011b}) performed numerical simulations  in which the spine and fan of a proper 3D null point are subjected to rotational and shear perturbations. They found that  rotations of the fan plane lead to current density accumulation about the spine,  and rotations about the spine lead to current sheets  in the fan plane. In addition,  shearing perturbations lead to 3D localised current sheets focused at the null point itself. Again, this is in good agreement with what we may expect for MHD wave behaviour from the 2D studies, i.e.  current accumulation at specific, predictable parts of the magnetic topology.

The first study of MHD wave behaviour in the neighbourhood of a fully 3D null point was investigated by McLaughlin et al. (\cite{MFH2008}). These authors examine the fast and Alfv\'en waves about both proper (i.e. 2.5D) and improper (fully 3D) null points. The authors utilise the WKB approximation to determine the transient properties of the modes in a linear, $\beta=0$ plasma regime. Their findings strongly suggest that the features of MHD waves about 3D nulls are not that different to the 2D results: the fast wave propagates across magnetic fieldlines according to the Alfv\'en-speed profile and the Alfv\'en wave is confined to magnetic field lines, leading to the waves accumulating in particular topological regions of the null point. However, their implementation of the WKB method is unable to address the question of whether modes couple in the fully 3D geometry, since their first-order WKB solution implicitly precludes the possibility and constrains the waves to see the magnetic field as locally uniform.

Thus, as it stands the question of what is the true behaviour of MHD waves in the neighbourhood of fully 3D magnetic null points remains unanswered. Specifically, what is the nature of the propagation and evolution of each MHD wave and, critically, what is the efficiency of mode-coupling and/or conversion due to the magnetic geometry and/or due to nonlinear effects? 


\begin{figure*}[t]
\centering
\includegraphics[width=17cm]{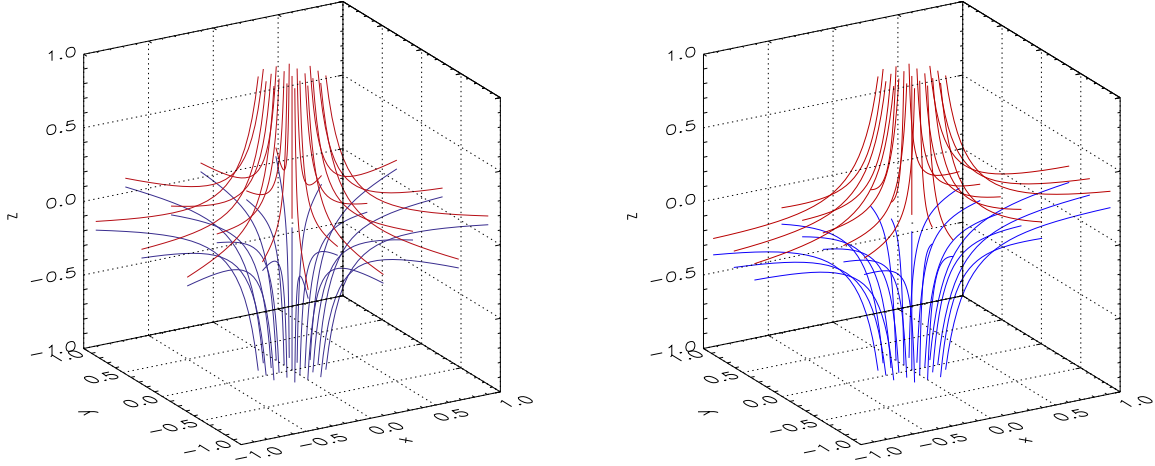}
\caption{Left: The azimuthally symmetric proper 3D null point ($\epsilon=1$). Right: An improper 3D null point where fieldlines are predominantly aligned with the $x$-axis ($\epsilon=0.5$).}
\label{fig1}
\end{figure*}

This paper investigates the behaviour of fast magnetoacoustic waves about fully 3D null points via numerical simulations and looks for evidence of coupling to the Alfv\'en mode in any form (e.g. linear, nonlinear and due to the 3D magnetic geometry). The paper is structured as follows: $\S \ref{section:2}$ outlines the specifics of the model and is subdivided into sections on the structure of the null points considered ($\S \ref{section:2.1}$), the mathematical model ($\S \ref{section:2.2}$), the method for isolating individual wave modes ($\S \ref{section:2.3}$), the specifics of the numerical solution ($\S \ref{section:2.4}$) and a brief outline of the supporting WKB solution ($\S \ref{section:2.5}$). $\S \ref{section:4}$ and $\ref{section:3.1}$ present the results of the simulations for the improper and proper radial null points respectively, and the conclusions are presented in $\S \ref{section:5}$.

\section{Mathematical Model}\label{section:2}

\subsection{Magnetic Null Points}\label{section:2.1}

In this paper, we investigate MHD wave behaviour about 3D potential magnetic null points. In Cartesian coordinates, these null points take the form:
\begin{eqnarray}
\mathbf{B}=\left[x,\epsilon y, -\left(\epsilon+1\right)z\right]  \label{eqn:3D_null_point}
\end{eqnarray}
where  the parameter $\epsilon$ is related to the eccentricity of the magnetic field lines, controlling the direction in which they predominantly align.  Parnell et al. (\cite{Parnell1996}) investigated and classified the different types of linear magnetic null points that can exist (our $\epsilon$ parameter is called $p$ in their work). For equation (\ref{eqn:3D_null_point}), the null point itself, i.e. ${\bf{B}}={\bf{0}}$, occurs at the origin.

Priest \& Titov (\cite{priesttitov}) identified two key topological features present near any 3D null point: the {\emph{spine}} and the {\emph{fan}} plane. The spine is an isolated fieldline that approaches, or leaves, the null point along the $z$-axis. In this paper, we only consider $\epsilon>0$ to restrict our attention to {\emph{positive}} null points, and as such the spine represents fieldlines approaching the null from above and below the $z=0$ plane. Thus, the $z=0$ plane, known as the fan, consists of radial field lines, confined to the $z=0$ plane and pointing radially away from the null point.

The eccentricity parameter $\epsilon$ alters the field topology as follows:
\def\labelitemi{$\bullet$}
\begin{itemize}
\item For $\epsilon=1$, the magnetic null point has azimuthal symmetry about its spine, with no prefered direction for fieldlines, and is known as a {\emph{proper}} null. Thus, as mentioned above, this is actually a 2.5D null point.
\item Null points which deviate from this cylindrical symmetry are known as {\emph{improper}} nulls. For $0\leq\epsilon\leq1$ the fieldlines curve to run parallel to the $x$-axis, and for $\epsilon\geq1$ curve to run parallel to the $y$-axis.
\item For $\epsilon=0$, we recover the simple 2D null point in the $xz-$plane, with a null line running through $x=z=0$.
\end{itemize}
See Parnell et al. (\cite{Parnell1996}) for more comprehensive information on the classification of different types of 3D null.

In this paper, we seek to determine the behaviour of fast MHD waves about potential 3D null points in the most general terms possible. However, analytical progress is unlikely when considering a completely general null point so we consider two different, specific null points that capture the range of topologies possible. The left panel of Figure \ref{fig1} shows the magnetic field line structure of our first choice: the proper $\epsilon=1$ null point (chosen for its azimuthally-symmetric fieldlines, which are (below) exploited in analytical approaches). The right hand panel shows the second: an improper null defined by $\epsilon=0.5$. Here fieldlines are predominantly aligned parallel to the $x-$axis, and by comparison we address the question of how MHD wave behaviour alters with the field topology and eccentricity.

\subsection{Governing MHD Equations} \label{section:2.2}

The three-dimensional, nonlinear, ideal, adiabatic MHD equations are solved numerically for a $\beta=0$ plasma. 
In most of the solar corona the plasma-$\beta$ is very low, ($\beta\ll1$, Gary \cite{Gary})  and so the 
$\beta=0$ approximation is arguably  valid. 
Here, we are specifically concerned with addressing whether the 3D magnetic field about realistic null points introduces behaviour absent in previous 2D models. Hence, in this study we utilise the $\beta=0$ approximation intentionally to prohibit the introduction of the  slow mode and restrict our attention to the behaviour of fast magnetoacoustic wave and its interplay with the Alfv\'en mode, on the understanding that the results presented are a first step towards understanding how waves behave at realistic null points in the corona; which are both 3D and have $\beta \neq 0$.

 Thus, the governing equations are as follows:
\begin{eqnarray}
\rho\left[\frac{\partial\mathbf{v}}{\partial t}+(\mathbf{v}\cdot\mathbf{\nabla})\mathbf{v}\right]&=& \left(\frac{\mathbf{\nabla}\times\mathbf{B}}{\mu} \right)\times\mathbf{B}\quad,\nonumber \\
\frac{\partial\mathbf{B}}{\partial t}&=&\mathbf{\nabla}\times(\mathbf{v}\times\mathbf{B})\quad,\nonumber\\
\frac{\partial\rho}{\partial t}&=& -\mathbf{\nabla}\cdot(\rho\mathbf{v})  \quad,\nonumber\\
\frac{\partial p}{\partial t} &=& - \mathbf{v}\cdot\nabla p-\gamma p \mathbf{\nabla}\cdot\mathbf{v}\quad.
\label{MHDeqns}
\end{eqnarray}
Here, standard MHD notation applies: $\mathbf{v}$ is plasma velocity, $p$ is plasma/thermal pressure, $\rho$ is density, $\mathbf{B}$ is the magnetic field/induction,  $\gamma=5/3$ is the adiabatic index, and $\mu$ is the magnetic permeability.

We consider an equilibrium state of $\rho=\rho_0$, $p=p_0$ (where $\rho_0$ and $p_0$ are constants), ${\mathbf{v}}=\mathbf{0}$ and  equilibrium magnetic field ${\bf{B}}=\mathbf{B}_0$. Finite, small perturbations of amplitude $\alpha \ll 1$ are considered in the form
$\rho=\rho_0 + \alpha\rho_1(\mathbf{r},t) $, $p=p_0 + \alpha p_1(\mathbf{r},t)$, $\mathbf{v}=\mathbf{0}+\alpha\mathbf{v}(\mathbf{r},t)$ and $\mathbf{B}=\mathbf{B}_0 +\alpha\mathbf{b}(\mathbf{r},t)$
and a subsequent nondimensionalisation using the substitution $\mathbf{v}=\overline{v}\mathbf{v}^*$,$\nabla={\nabla^*}/{L}$, $\mathbf{B}_{0}=B_{0}\mathbf{B}_{0}^{*}$, $\mathbf{b}=B_{0}\mathbf{b}^{*}$, $\mathbf{t}=\overline{t}\mathbf{t}^{*}$, $p_1=p_{0}p_{1}^{*}$ and $\rho_1=\rho_{0}\rho_{1}^{*}$ is performed, with the additional choices $\overline{v}={L} / {\overline{t}}$ and $\overline{v}={B_{0}} / {\sqrt{\mu\rho_{0}}}$.  The resulting nondimensionalised, governing equations of the perturbed system are:
\begin{eqnarray}
\frac{\partial \mathbf{v}}{\partial t} &=& \left(\nabla\times\mathbf{b}\right)\times\mathbf{B}_0 + \mathbf{N}_{1} \nonumber\\
\frac{\partial\mathbf{b}}{\partial t} &=& \nabla\times\left( \mathbf{v} \times\mathbf{B}_{0}\right) +  \mathbf{N}_{2} \nonumber\\
\frac{\partial \rho_1}{\partial t} &=& -\nabla\cdot \mathbf{v} + {N}_{3}\nonumber\\
\frac{\partial p_1}{\partial t} &=& -\gamma \nabla \cdot \mathbf{v} + {N}_{4} \nonumber\\
{\bf{N}}_1 &=& \left(\nabla\times\mathbf{b}\right)\times\mathbf{b} -\rho_{1}\frac{\partial \mathbf{v}}{\partial t}  - \left(\mathbf{v}\cdot\nabla\right)\mathbf{v} \nonumber \\
{\bf{N}}_{2} &=& \nabla \times \left(\mathbf{v}\times\mathbf{b}\right)\nonumber \\
{{N}}_3 &=& -\nabla \cdot \left(\rho_{1} \mathbf{v} \right)  \nonumber \\
{{N}}_4 &=& \mathbf{v} \cdot \left(\nabla p_1 \right) -\gamma p_{1} \left(\nabla\cdot\mathbf{v}\right)\label{equation_MHD}
\end{eqnarray}
where terms ${\bf{N}}_i$ are the nonlinear components correct to $\mathcal{O}(\alpha^2)$. The star indices have been dropped, henceforth all equations are presented in a nondimensional form.
The  equations are merged into one governing PDE: 
\begin{eqnarray}
\frac{\partial^2 \mathbf{v}}{\partial t^2}&=&\left\lbrace\mathbf{\nabla}\times\left[\mathbf{\nabla}\times\left(\mathbf{v}\times\mathbf{B}_0\right)\right]\right\rbrace\times\mathbf{B}_0 + \mathbf{N} \nonumber \\
\mathbf{N}&=&\left\lbrace\mathbf{\nabla}\times\left[\mathbf{\nabla}\times\left(\mathbf{v}\times\mathbf{b}\right)\right]\right\rbrace\times\mathbf{B}_0
\nonumber \\
& & + \left\lbrace\mathbf{\nabla}\times\left[\mathbf{\nabla}\times\left(\mathbf{v}\times\mathbf{b}\right)\right]\right\rbrace\times\mathbf{b}
+\left(\nabla\times\mathbf{b}\right)\times    \left[\nabla\times\left(\mathbf{v}\times\mathbf{B}_{0}\right)\right] \nonumber \\
& & + \left(\mathbf{\nabla}\cdot\mathbf{v} - \mathbf{v} \cdot  \mathbf{\nabla}\right)\left(\mathbf{\nabla}\times\mathbf{b}\right)\times\mathbf{B}_0 \nonumber\\
& & - \rho_1 \left\lbrace\mathbf{\nabla}\times\left[\mathbf{\nabla}\times\left(\mathbf{v}\times\mathbf{B}_0\right)\right]\right\rbrace\times\mathbf{B}_0 \nonumber \\
& & -  \left[ \left(\mathbf{\nabla}\times\mathbf{b}\right)\times\mathbf{B}_0  \cdot\mathbf{\nabla}\right] \mathbf{v}
  \label{equation_9}
\end{eqnarray}
The first term dominates the linear regime of  $\mathcal{O}\left(\alpha\right)$ and the terms  $\mathbf{N}$ represent nonlinear terms of  $\mathcal{O}(\alpha^2)$.

This study utilises numerical solutions of the full nonlinear MHD equations (\ref{MHDeqns}), using the numerical method detailed in $\S \ref{section:2.4}$. In conjunction, analytical treatment of  wave equation (\ref{equation_9}) derived via second-order perturbation theory is used to create a special coordinate system that will isolate and identify individual  wave modes ($\S \ref{section:2.3}$), create a WKB approximation ($\S \ref{section:2.5}$) and to seek analytical confirmation of observed phenomena in the numerical simulations where possible ($\S \ref{section:A}$).

\subsection{Isolating MHD modes}\label{section:2.3}

This paper  investigates the transient MHD mode behaviour  about a general 3D null point as well as the possible mode coupling/conversion that may occur. To do so, it is necessary to construct a novel coordinate system that allows individual, pure MHD wave modes to be generated, identified and tracked. In turn, such distinction requires a clear definition of the possible modes that can occur.

As outlined in  $\S \ref{section:1}$, the concept and terminology of the three classes of MHD wave  (Alfv\'en, fast and slow) that originated within a homogenous MHD plasma description still remains valid and useful in describing inhomogenous media, although distinguishing between the individual modes can become increasingly difficult depending upon the exact nature of the inhomogenity. The  $\beta=0$, uniform density, 3D magnetic null points under consideration in this paper are very specific examples of plasma inhomogenity, namely that of a nonuniform  magnetic topology. Hence, the behaviour of the two linear waves sustainable here (Alfv\'en and fast) is solely due to the Lorentz force, and as such we find that the most appropriate and useful definitions of, and distinctions between, the two modes are purely in these terms.

\begin{figure}
\resizebox{\hsize}{!}{\includegraphics{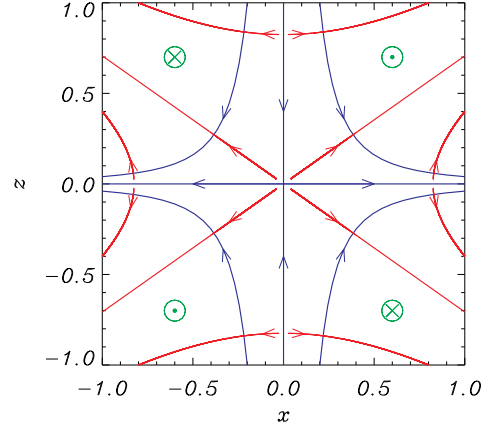}}
\caption{Fieldlines of magnetic induction $\mathbf{B}_0$ (blue) and the field  $\mathbf{C}$ (red) perpendicular to both the magnetic induction and the flux function, in the case $\epsilon=1$ for the $y=0$ plane. Due to azimuthal symmetry other such planes have identical fieldline structure. The flux function $\mathbf{A}$ (green) in this case is azimuthal about the spine, running antiparallel to $\hat{\vec{\theta}}$ above the fan plane and parallel to $\hat{\vec{\theta}}$ below ($\otimes$ corresponds to a vector directed towards the point of view, and $\odot$ away). These vector fields are everywhere perpendicular thus suitable for forming an orthogonal coordinate system, although fields $\mathbf{C}$ and  $\mathbf{A}$ are undefined on the spine and fan plane.}
\label{fig2}
\end{figure}

The nondimensionalised Lorentz force:
\begin{equation}
\mathbf{F}=\left(\mathbf{\nabla}\times\mathbf{B}\right)\times\mathbf{B}=(\mathbf{B}\cdot\mathbf{\nabla})\mathbf{B}-\frac{\nabla}{2}(\mathbf{B}\cdot\mathbf{B})\label{eq:Lforce}
\end{equation} 
is the sole driving force in our $\beta=0$ system, and is denoted $\mathbf{F}$. The first term of the right hand side of the equation (\ref{eq:Lforce}) can be interpreted as magnetic tension, and the second term is a gradient in magnetic pressure. We note that the force has no term parallel to the magnetic induction (${{\mathbf{F}\cdot\mathbf{B}}} =0$) and it follows that no $\beta=0$ MHD wave can be driven by or associated with perturbations in this direction, {\emph{i.e.}} they are all perpendicular perturbations relative to total magnetic field. Thus, we define our two waves as follows:
\begin{itemize}
\item[1)] {The (linear) Alfv\'en wave is driven only by magnetic tension. This results in a wave that propagates anisotropically as a transverse wave, confined to and guided along the magnetic fieldlines.} 
\item[2)] {The (linear) fast magnetoacoustic wave is driven both by magnetic tension and magnetic-pressure gradients. It propagates isotropically via a mixture of transverse and longitudinal motions (due to both magnetic tension and magnetic-pressure gradient respectively).}
\end{itemize}
These definitions are used for the specific purpose of creating a special orthogonal coordinate system that can be used to distinguish between these two modes. Note that this will not be valid along the spine or along the fan plane, as here the waves are degenerate: at these locations both the fast wave and Alfv\'en wave propagate only as transverse, tension-driven waves. Elsewhere, such a coordinate system can be obtained by exploiting a mathematically-invariant direction.

For instance, if we consider initial perturbations to the equilibrium magnetic field ($\hat{\mathbf{B}}_0$) in an invariant direction denoted $\hat{\mathbf{e}}_i$, then due to $\partial / \partial \hat{\mathbf{e}}_i = 0 $, the magnetic-pressure gradient is eliminated. As such, this direction is inextricably linked with the Alfv\'en mode as per our definition, i.e. perturbations in this direction give rise to waves driven only by magnetic tension that propagate as transverse waves along magnetic field lines. The existence of an invariant direction is just as much a physical requirement for the existence of Alfv\'en waves as it is a convenient mathematical method for isolating them. Various authors have stated that in the absence  of such an invariant coordinate, only the magnetoacoustic waves can exist, and hence only certain special topologies permit the Alfv\'en mode (see, e.g., Parker \cite{parker91}; Van Doorsselaere et al. \cite{Tom2008}; Goossens et al. \cite{goosens2011}).

Alternatively, we can consider perturbations in a direction that is perpendicular to both $\mathbf{B}_0$ and $\hat{\mathbf{e}}_i$. Here both magnetic pressure and magnetic tension can be present and thus we assume this direction is associated with the fast mode as per our definition. That is to say,   both magnetic tension and magnetic-pressure gradients sustain an isotropic wave  that propagates via longitudinal and transverse motions.

To reiterate: in this paper we aim to construct a special coordinate system, for the 3D null, of the form  $\hat{\mathbf{B}}_0$,    $\hat{\mathbf{e}}_i$ and  $\hat{\mathbf{e}}_{i} \times \hat{\mathbf{B}}_0$, such that motions in the $\hat{\mathbf{e}}_i$-direction correspond to the Alfv\'en mode and that motions in the $\hat{\mathbf{e}}_{i}   \times  \hat{\mathbf{B}}_0$-direction correspond to fast mode.

Similar special coordiante systems have been previously utilised  in various 2D MHD models, an example of which  is  the coordinate system used by McLaughlin \& Hood (\cite{MH2004}; \cite{MH2005}; \cite{MH2006a}) to drive pure fast and Alfv\'en modes at a 2D X-point. In fact, the essence of any 2D mathematical model is to impose an obvious invariant direction, and thus facilitating the creation of such a coordinate system. However, for a fully 3D null point ($\epsilon \neq 0, \epsilon \neq 1$), there is no obvious invariant direction and, unlike 2D models, there is no clear way to discern which of the directions perpendicular to $\mathbf{B}_0$  would be associated with unique MHD wave modes.

We look for clues in the proper ($\epsilon=1$) null case, which has obvious azimuthal symmetry. Considering the proper null point in cylindrical polar coordinates reduces the problem to 2.5D. We first seek to find the coordinate system which corresponds to unique wave modes in this specific case, then look for a generalised system (of which this would be a specific case) that would work with any 3D null point.

The equilibrium magnetic field with $\epsilon=1$ rephrased in cylindrical polars is $\mathbf{B}_{0}=\left[r,0,-2z\right]$
where $r^2=x^2+y^2$. Under our definition, the Alfv\'en wave will be associated with the invariant direction $\hat{\vec{\theta}}$. Thus, it follows that the fast wave is  associated with the direction that completes the orthogonal set, i.e. $\hat{\vec{\theta}}\times\mathbf{B}_0$. Considering the  linear components of the wave equation (\ref{equation_9}) parallel to the orthogonal vectors  $\hat{\vec{\theta}}$, $\mathbf{B}_0$ and $\hat{\vec{\theta}}\times\mathbf{B}_0$,   in cylindrical polars for $\epsilon=1$ leads to three decoupled wave equations governing the linear Alfv\'en, fast and slow (absent) mode respectively:

\begin{eqnarray}
\displaystyle \frac{\partial^{2}\mathbf{v}}{\partial t^2}\cdot\hat{\vec{\theta}} &=& \left[-1 + r \frac{\partial}{\partial r} +r^{2}\frac{\partial^2}{\partial r^2}+ 4z \frac{\partial}{\partial z} +4z^{2} \frac{\partial^2}{\partial z^2} \right. \nonumber \\
& & \left. - 4rz\frac{\partial^2}{\partial r \partial z}\right] v_\theta \nonumber \\
\frac{\partial^{2}\mathbf{v}}{\partial t^2}\cdot\left(\hat{\vec{\theta}}\times\mathbf{B}_0\right) &=& -\left(r^{2}+4z^{2}\right)\left[\left(4 \frac{\partial}{\partial z}+2z \frac{\partial^2}{\partial z^2}-\frac{2z}{r^2}+\frac{2z}{r}\frac{\partial}{\partial r} \right. \right. \nonumber \\
& &  + \left. \left. 2z \frac{\partial^2}{\partial r^2}\right)v_{r} \right.   + \left. \left(3\frac{\partial}{\partial r} + r\frac{\partial^2}{\partial r^2} + r\frac{\partial^2}{\partial z^2} \right) v_{z}\right]  \nonumber \\
 \frac{\partial^{2}\mathbf{v}}{\partial t^2}\cdot\mathbf{B}_{0} &=&  0 
\label{eq:lindecoupledwave}
\end{eqnarray}

This treatment offers a clear demonstration of (linearly) decoupled waves for this specific case and provides confirmation that perturbations in $\hat{\vec{\theta}}$ will be associated with the Alfv\'en wave and perturbations in $\hat{\vec{\theta}}\times\mathbf{B}_0$ with the fast wave. Such a treatment in cylindrical polars has been used in previous studies of proper magnetic null points, for example Galsgaard et al. (\cite{Galsgaard2003}) utilised a symmetric helical motion (in our terminology, a perturbation in $\hat{\vec{\theta}}$, to study an Alfv\'en wave propagating about a proper, $\epsilon=1$ null point). However, it is not obvious how this method would transfer to the more general improper null case.

\subsubsection{Special coordinate system: ${\bf{A}}$, ${\bf{B}}_0$ and ${\bf{C}} = {\bf{A}} \times {\bf{B}}_0$}\label{section:2.3.1}

The flux function/magnetic vector potential is defined such that $\mathbf{B}_0=\nabla\times\mathbf{A}$. For any simple, single potential null set-up, the magnetic helicity is zero (i.e. $\mathbf{A}\cdot\mathbf{B}_{0}=0$) and hence ${\bf{A}}$ is aligned with the invariant direction.  Thus, we hypothesise that about potential null points the (linear) Alfv\'en wave is always associated with the flux function direction. For the $\epsilon=1$ null, we find the flux function, and thus the special coordinate system, would be (in cylindrical polars):
\begin{equation}
\begin{array}{l}
\displaystyle \mathbf{B}_0=\left[r,0,-2z\right]\\
\displaystyle \mathbf{A}=\left[0,-rz,0\right]  \propto  \hat{\vec{\theta}}\\
\displaystyle \mathbf{C}=\left[2rz^2,0,zr^{2}\right]   \propto  (\hat{\vec{\theta}}\times\mathbf{B}_{0})
\end{array} 
\end{equation}
and we confirm that the vectors form an orthogonal set $\mathbf{A}\cdot\mathbf{B}_0=\mathbf{A}\cdot\mathbf{C}=\mathbf{B}_{0}\cdot\mathbf{C} = 0$. 
Thus, to generate and isolate a pure Alfv\'en mode, we consider perturbations in the $\hat{\mathbf{A}}$-direction only, and similarly for the pure fast wave, we must consider perturbations in the direction $\hat{\mathbf{C}}$, where $\mathbf{C}=\mathbf{A}\times\mathbf{B}_0$.  The system for $\epsilon=1$ can be considered a special case of cylindrical polars that changes about the fan plane, and is sufficiently similar to be consistent with the decoupled linearised wave equations for the cylindrical polars (\ref{eq:lindecoupledwave}). Figure \ref{fig2} illustrates the vector fields for the $\epsilon=1$ case.

By determining the flux function for general $\epsilon$, we extend this to create  an orthogonal, curvilinear coordinate system  that is  based on the equilibrium magnetic field  (${\bf{B}}_0$), the flux function (${\bf{A}}$) and their cross product  (${\bf{C}}= {\bf{A}}\times {\bf{B}}_0$). These general vector fields are found to take the Cartesian form:
\begin{equation}
\begin{array}{l}
\displaystyle {\mathbf{B}}_{0}=\left[x,\epsilon y,-(\epsilon+1)z\right]\\
\displaystyle \mathbf{A}=\left[zy,-\epsilon xz,(1-\epsilon)xy\right]\\
\displaystyle \mathbf{C}=\left[C_x,C_y,C_z\right] \\
\displaystyle C_x = x \left[ \left(\epsilon^2-\epsilon\right)y^{2}+\left(\epsilon+1\right)z^{2}\right] \\
\displaystyle C_y = y  \left[ \left(1-\epsilon \right)x^{2}+\left(\epsilon+1\right)z^{2}\right] \\
\displaystyle C_z = \epsilon z \left(x^{2}+y^{2}\right)
\end{array} 
\label{eq:gencoordsyst}
\end{equation}
with unit normals $\mathbf{\hat{B}}=\mathbf{B}/|\mathbf{B}|$, $\mathbf{\hat{A}}=\mathbf{A}/|\mathbf{A}|$ and $\mathbf{\hat{C}}=\mathbf{C}/|\mathbf{C}|$. Note that on the line of the spine ($x=y=0$) and the fan plane ($z=0$), the identity $\mathbf{B} = \nabla \times \mathbf{A}$ no longer holds, rendering the system locally invalid. As such, {\emph{the spine or fan  cannot be used to drive pure modes}} under this coordinate system. However, there is a good physical explanation for this, due to a degeneracy between the fast and Alfv\'en modes on the fan plane and spine, i.e. a wave propagating along the spine or along one of the fieldlines on the fan plane is driven by tension only, as a transverse oscillation, and so along these specific/particular fieldlines the  fast mode is degenerate with  the Alfv\'en mode.

This special coordinate system  (equation \ref{eq:gencoordsyst}) is at the heart of the work presented in this paper. It is used in the following manner: Initially, it is employed to introduce a pure, linear,  fast magnetoacoustic wave at a computational boundary by driving a velocity pulse in the direction $\hat{\mathbf{C}}$. Numerical experimentation ($\S \ref{section:4}$ \& $\ref{section:3.1}$) confirm that this special orthogonal coordinate system successfully isolates the fast mode. The coordinate system is then used to seek evidence of linear and/or nonlinear coupling to the Alfv\'en mode, i.e. if we drive a pure fast wave ($\mathbf{v}_{\hat{\mathbf{C}}} \neq 0, \mathbf{v}_{\hat{\mathbf{A}}} = 0$) and mode conversion occurs then under our coordinate system velocity components in the direction $\hat{\mathbf{A}}$ will become manifest.

\subsection{Numerical Solution}\label{section:2.4}

Nondimensionalised versions of the governing nonlinear MHD equations (\ref{equation_MHD}) are solved using the {\emph{LARE3D}} numerical code (a Lagrangian-Eulerian remap scheme for MHD, see Arber et al. \cite{LAREpaper})  with magnetic equilibria corresponding to both proper and improper 3D null points. Each scenario introduces a pure, planar fast magnetoacoustic wave pulse along the upper $z$-boundary, by driving the following sinusoidal profile: 
\begin{equation}
\mathbf{v_{\hat{C}}} = \alpha \sin{(2\pi t)} \;\;, \quad  \mathbf{v_{\hat{A}}} =0\;\;, \quad \mathbf{v}_{\mathbf{\hat{B}}_0}=0  \quad \mathrm{for} \quad 0 \le t \le 0.5
\label{pulse}
\end{equation}
 Numerical tests determined that the amplitude $\alpha=0.001$ was sufficiently small relative to other parameters to generate a linear wave and for the rest of this paper we set $\alpha=0.001$. The other boundary conditions are set as zero-gradient conditions. The simulations  utilise a uniform numerical grid with  domain $-1 \le x \le 1$, $-1 \le y \le 1$, $-1 \le z \le 2$ and $720 \times 720 \times 1080$ grid points, giving an effective resolution of $\delta x \approx \delta y \approx \delta z \approx {1/360}  $. The results presented focus into the region $-1 \le x \le 1$, $-1 \le y \le 1$, $-1 \le z \le 1$, i.e. a subset of the full numerical domain.

Since ${\mathbf{{\hat{C}}}}$ is undefined along the spine and the fan, driving ${\mathbf{v_{\hat{C}}}}$ generates a planar wave with a small hole about the spine (for the top-boundary  driven simulations) or a gap along the fan (for the side-driven simulations). This is essential to avoid the mode degeneracy on the spine and/or fan-plane fieldlines.

After the near-planar pulse is generated, a small trailing wake is also introduced  due to dispersion (e.g. Uralov \cite{uralov}). Such a wake does not impact upon the dynamics of the main/lead pulse.

\subsection{WKB Approximation}\label{section:2.5}

Semi-analytical WKB approximations of fast waves about both proper and improper  null points  were previously considered by McLaughlin et al. (\cite{MFH2008}). Succinctly,  the WKB method is an asymptotic series approximation  for low-frequency wave forms and has previously proven useful in understanding  null point models. The WKB solution for fast waves is presented in $\ref{section:4}$ overlaying our numerical results. More information regarding the WKB method (Wentzel \cite{W}; Kramers \cite{K}; Brillouin \cite{B}) can be found in Bender \& Orszag (\cite{BendO}).

However, McLaughlin et al. (\cite{MFH2008}) only consider the first-order WKB approximation, and thus  their WKB implementation cannot account for the possibility of mode coupling, which is one of the chief concerns of this paper. As such,  comparison to the WKB method serves a two-fold purpose: if the solutions disagree it indicate possible occurrences of mode coupling or, alternatively, it indicates numerical inaccuracies. In addition, due to our special coordinate system being undefined on the spine and fan, the near-planar waves presented in the numerical solutions have the aforementioned gap around the spine. However, the mathematical nature of the WKB solution allows one to avoid the question of whether it is possible to drive a pure mode on the spine or fan and simply determines which path a fluid element  would take if it were possible. As such, discrepancies between the WKB solution and numerical solutions on the spine or fan plane are entirely to be expected and indicative of neither coupling nor numerical error.

\section{Improper Null}\label{section:4}
\begin{figure*}
\sidecaption
\includegraphics[width=12cm]{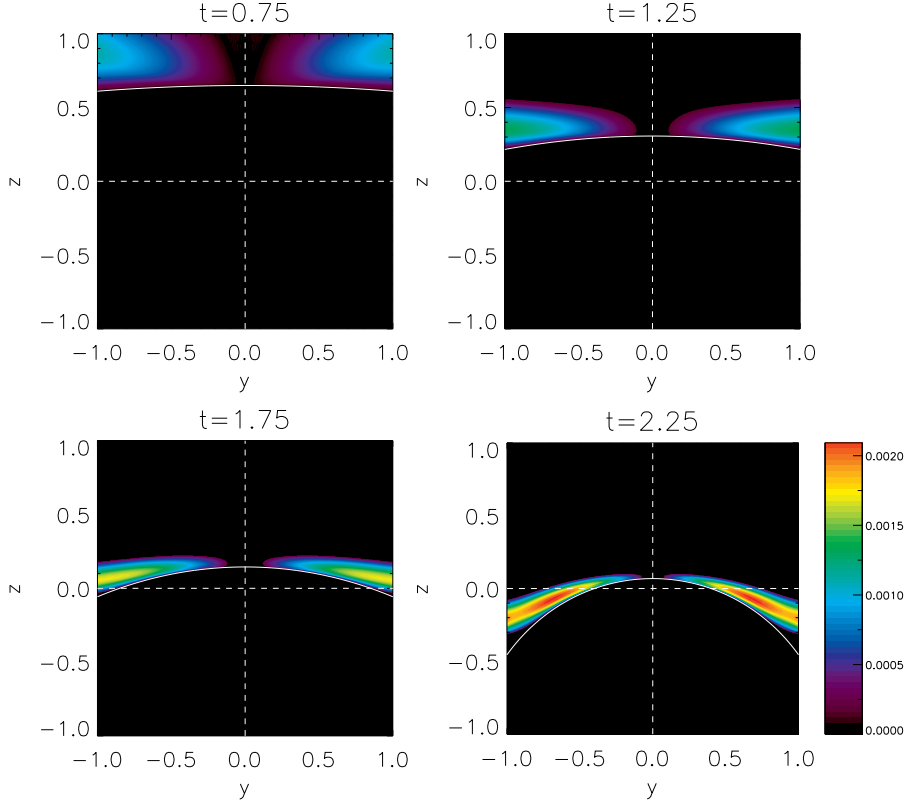}
\caption{Contour plots illustrating the propagation of $|\mathbf{v}|$ about the improper 3D null point for the $x=0$, $yz$-plane. The position of the lead wave front according to the WKB solution shown in white, which is in good agreement with the numerical results (sufficiently far from the boundary).}
\label{fig6}
\end{figure*}

\begin{figure*}
\sidecaption
\includegraphics[width=12cm]{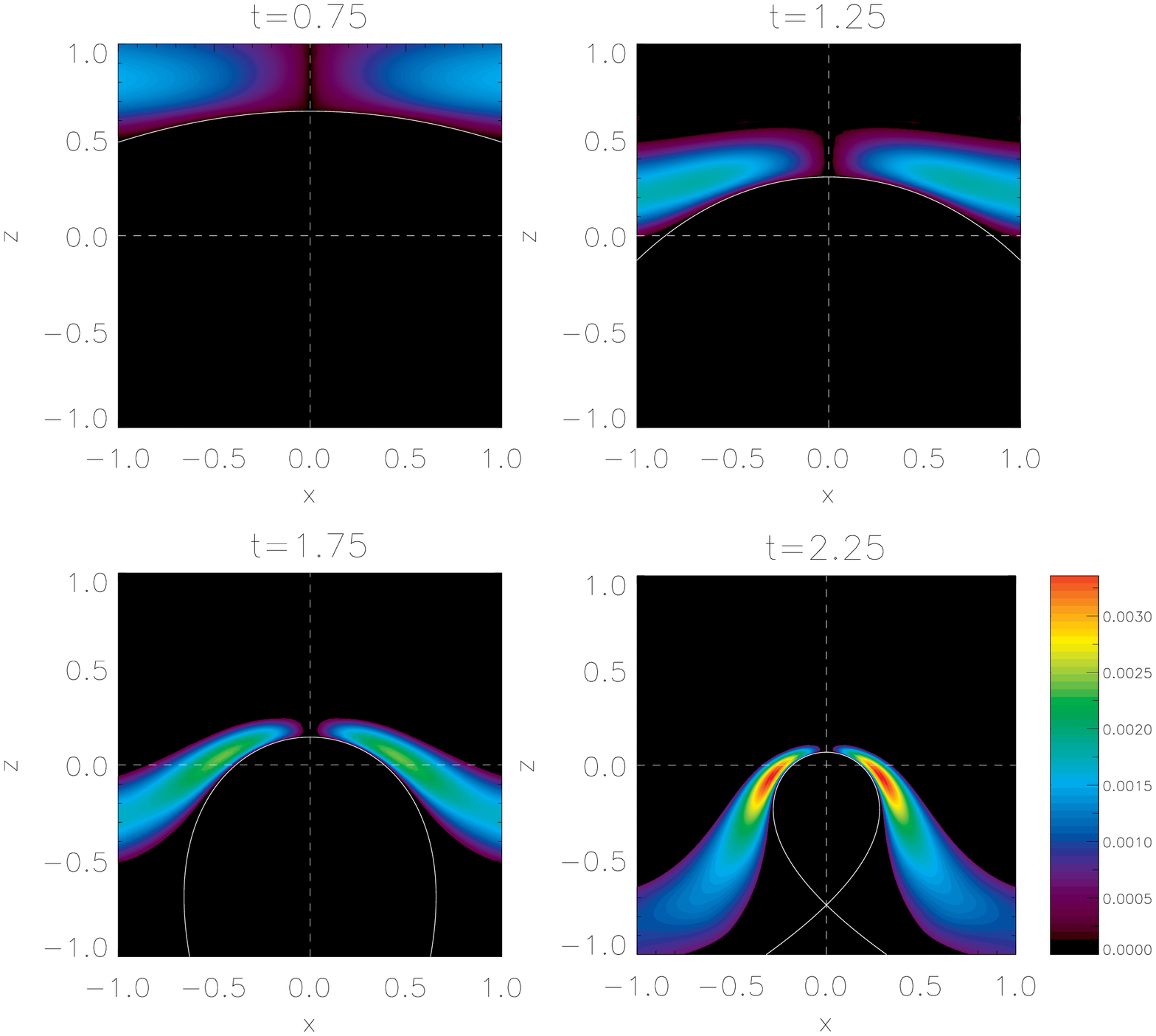}
\caption{Contour plots illustrating the propagation of $|\mathbf{v}|$ about the improper 3D null point for the $y=0$, $xz$-plane. The position of the lead wave front according to the WKB solution shown in white, which is in good agreement with the numerical results (sufficiently far from the boundary).}
\label{fig7}
\end{figure*}

Let us first consider the improper radial null point, in the specific case of $\epsilon=0.5$, where a velocity pulse (equation \ref{pulse}) is driven in  $\mathbf{v}_{\hat{\mathbf{C}}}$ along the top computational boundary (i.e. $z=2$). The resultant propagation  is seen in figures \ref{fig6} and \ref{fig7}, which figures show $|\mathbf{v}|$ in  the planes defined by both $x=0$ and $y=0$ respectively.

\emph{In both figures \ref{fig6} and \ref{fig7} we observe a refraction effect which focuses the wave towards the null point}. This refraction/wrapping effect has previously been identified as a key  feature of the behaviour of the fast magnetoacoustic mode about 2D nulls (e.g. McLaughlin \& Hood \cite{MH2004}) and is dictated by the Alfv\'en-speed profile around the null point, namely that the wave propagates from regions of high to low Alfv\'en speed. 

Here, the $x=0$ plane (Figure \ref{fig6}) shows the wave propagating in the region with the \lq{gentlest}\rq{ } Alfv\'en-speed profile: $c_{A}^2= \frac{1}{4}\left(y^{2}+9z^{2}\right)$ and the $y=0$ plane (Figure \ref{fig7}) shows the wave propagating in the region with the strongest Alfv\'en-speed profile: $c_{A}^2= \frac{1}{4}\left(4x^{2}+9z^{2}\right)$. This difference in Alfv\'en-speed profile directly corresponds to the difference of the rate of refraction in these figures, i.e. we observe the $y=0$ plane (Figure \ref{fig7}) to \lq\lq{wrap}\rq\rq{} around the null point more readily than that in the $x=0$ plane (Figure \ref{fig6}).

The WKB solution for the leading edge of a planar fast wave is also shown in Figure \ref{fig6} and \ref{fig7} (in white). The agreement between the leading edge of the wave front observed in $|\mathbf{v}|$ and the WKB solution near the null point is very good: the (numerical) wave experiences refraction along the same profile and at the same rate as prescribed by the analytical approximation. Towards the boundaries the numerical front increasingly lags behind the WKB front as time evolves: this not a physical attribute of the fast wave but rather a consequence of the boundary conditions.  However, this does not affect the key behaviour close to the null where the agreement between the two solutions is good.  There is a further discrepancy where the WKB solution exists across the spine whereas the pulse does not. However, this is expected as we purposefully did not drive any component along the spine due to the mode degeneracy discussed in $\S \ref{section:2.4}$.

The agreement between the numerical solution and  the WKB solution (which itself precludes mode coupling)  suggests that the pulse is propagating only as the fast magnetoacoustic wave  and thus no conversion to the Alfv\'en mode occurs. To verify this, we now utilise the special coordinate system to inspect the individual  constituent modes of the pulse. 

\subsection{${\hat{\mathbf{C}}}$: proxy for the fast mode}\label{section:4.1}

We now compare $|\mathbf{v}|$ and $|{\mathbf{v}}_{\hat{\mathbf{C}}}|$; \emph{viz.} the whole disturbance and the constituent fast mode. We find that the spatial extent, i.e. positions of the leading, middle and trailing wavefronts, of  $|\mathbf{v}_{\hat{\mathbf{C}}}|$ is identical to that of $|\mathbf{v}|$ at any given time, and that the difference in magnitude between the two is $\mathcal{O}(\alpha^2)$. 
Hence, we observe its (linear) transient behaviour about a proper null point to be dominated by the refraction effect, focusing the pulse at the null point. As the difference between $|\mathbf{v}|$ and $|\mathbf{v}_{\hat{\mathbf{C}}}|$ is non-zero, it is clear that nonlinear effects, possibly $\mathcal{O}(\alpha^2)$ mode conversion, are present but the two are linearly equivalent. We now investigate  the other orthogonal directions.

\subsection{${\hat{\mathbf{A}}}$: proxy for the Alfv\'en mode}\label{section:4.2}
By considering $|\mathbf{v}_{\hat{\mathbf{A}}}|$, we find that there is no velocity perturbation in this direction at any location or time in the simulation. Hence, no figure is presented, since  $|\mathbf{v}_{\hat{\mathbf{A}}}|$ is identically zero for all time. As detailed in $\S \ref{section:2.3}$, this velocity component is in the direction of the Alfv\'en wave, and as such the results indicate that {\emph{a pure fast wave propagating about an improper 3D null does not generate an Alfv\'en wave due to either linear or nonlinear coupling}}.

\subsection{${\hat{\mathbf{B}}}_0$: proxy for the field-aligned motions}\label{section:4.3}

\begin{figure*}
\sidecaption
\includegraphics[width=12cm]{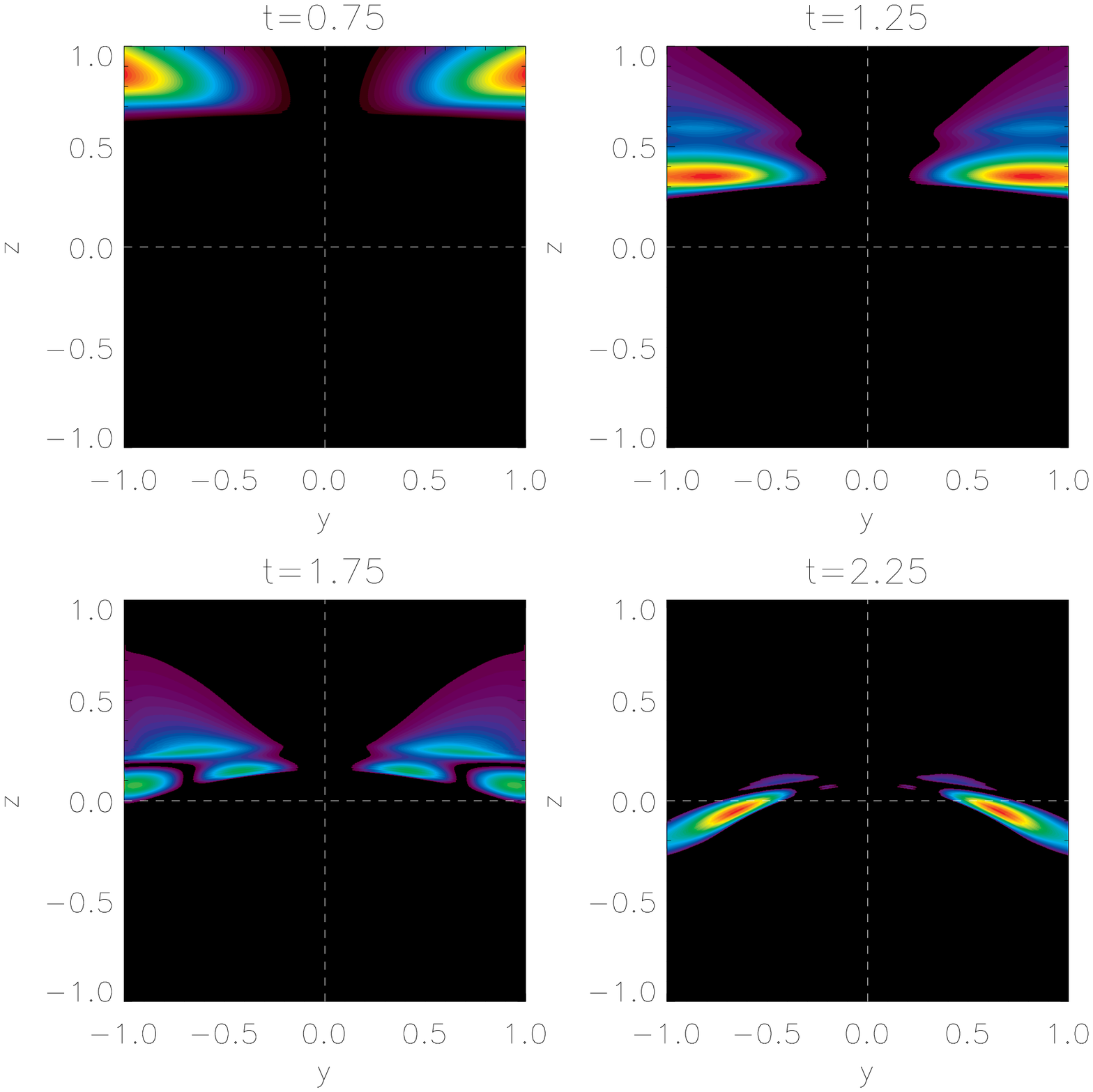}
\caption{The nonlinear velocity component  $|\mathbf{v}_{\hat{\mathbf{B}}_0}|$ for the $x=0$ cut of the improper null point: a small disturbance is created that mostly occupies the same geometry as $|\mathbf{v_{\hat{C}}}|$. Each contour is scaled relevant to the amplitude at the specific times to enhance the contrast of the features.}
\label{fig9}
\end{figure*}

\begin{figure*}
\sidecaption
\includegraphics[width=12cm]{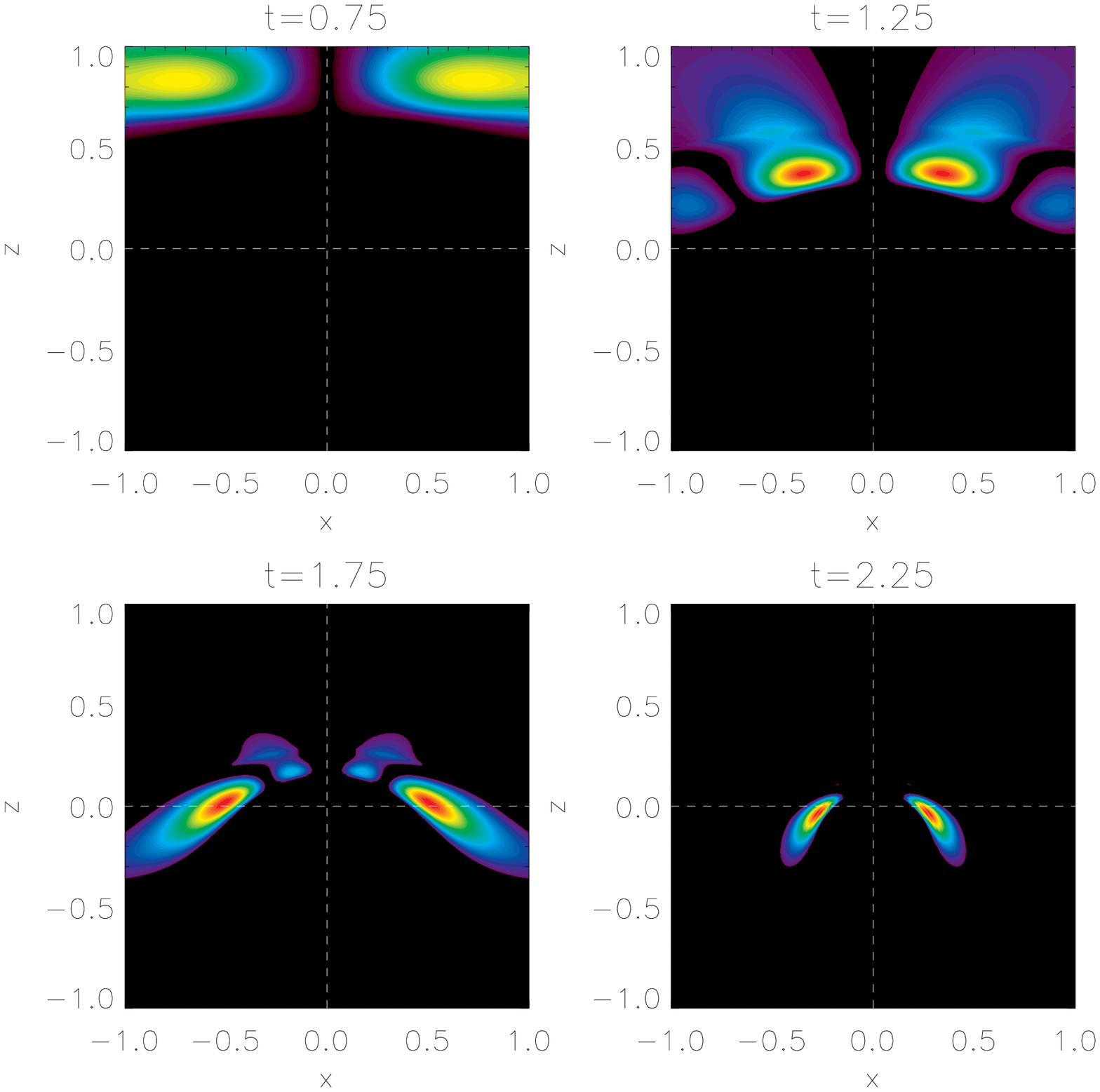}
\caption{The nonlinear velocity component  $|\mathbf{v}_{\hat{\mathbf{B}}_0}|$ as per Figure \ref{fig9} corresponding to the $y=0$ cut.}
\label{fig10}
\end{figure*}

Figures \ref{fig9} and \ref{fig10} show the velocity component in  $|\mathbf{v}_{\hat{\mathbf{B}}_0}|$  (motions aligned with the equilibrium magnetic field) over the same time scale as Figures \ref{fig6} and \ref{fig7}. We see that the wave shown is a nonlinear disturbance of $\mathcal{O}(\alpha^2)$ that, bar a small trailing wake, is confined to within the same spatial locations as the linear fast wave. 
However, this nonlinear disturbance is  of a different geometry to that of  the fast wave  and  splits into multiple peaks on the approach to the fan plane. These multiple peaks appear to be akin to the lobes seen in McLaughlin \& Hood (\cite{MH2006b}). During the evolution, we find that the disturbance of  $\mathcal{O}(\alpha^2)$ is entirely sustained by and dependent upon the propagating fast pulse of  $\mathcal{O}(\alpha)$,  which  acts as a {\emph{progenitor wave}} that creates this {\emph{daughter disturbance}} instantaneously as it propagates.

As the fast wave propagates about the proper null point, it causes small disturbances to the equilibrium (which can be thought of as both a force free magnetic field and a convective/flow free fluid) as it passes. This effect is the subsequent field/fluid response (as a consequence of frozen flux the two are unsurprisingly connected) that settles to the initial configuration. Due to this dependency on the propagating fast mode (it eventually returns to zero once the fast wave has passed an area, as opposed to being generated but then propagating independently) we have used the term \textit{disturbance} when describing this effect as opposed to \textit{wave} or \textit{mode}. The effect responsible for the nonlinear disturbance is related (but not identical to) what has been called the \textit{ponderomotive} effect in MHD literature (see e.g. Verwichte \cite{erwin1999a}; Verwichte et al. \cite{erwin1999b};   McLaughlin et al. \cite{james2011b}). As this clearly has had no effect on the mode conversion process (which is absent) this effect is not further investigated.

Note, the numerical experiment was repeated
with the driving of a pure fast wave along one of the faces perpendicular
to the fan plane, i.e. a \lq{side-driven}\rq{} fast wave. The
key results were found to be the same as those of the top-driven
simulation, confirming that the coupling and transient behaviour
is independent of spine/plane geometry.

\section{Proper Null, $\epsilon=1$}\label{section:3.1}
The numerical experiments were also considered for the proper null point ($\epsilon=1$). Here, the results where found to be consistent with, and analogous to, those in $\S \ref{section:4}$, namely that:
\begin{itemize}
\item[1)] {The whole pulse and the fast wave component, $|\mathbf{v}|$ and $|{\mathbf{v}}_{\hat{\mathbf{C}}}|$, are spatially identical and linearly equivalent, with a small difference of $\mathcal{O}(\alpha^2)$. In both, we find a pulse that refracts along the Alfv\'en-speed profile, in good agreement with WKB solution.} 
\item[2)] {There is no coupling, linear or nonlinear, to the Alfv\'en mode ($|\mathbf{v_{\hat{A}}}|$ is permanently zero throughout).}
\item[3)] {There propagating fast wave sustains an instantaneous, nonlinear disturbance in $|\mathbf{v}_{\hat{\mathbf{B}}_0}|$.}
\end{itemize}
Additionally, as the proper null point assumes azimuthal symmetry analytical confirmations of mode decoupling can be obtained (see appendix \ref{section:A}). The results only differ in the rates of refraction and time for accumulation at the null point, as the proper null has a different Alfv\'en-speed profile.

\section{Conclusion}\label{section:5}

We have studied both the transient properties of fast magnetoacoustic waves and the nature of linear and nonlinear mode coupling in the vicinity of two potential 3D null points: the proper ($\epsilon=1$) null, which can be treated as a 2.5D problem using cylindrical polars, and the fully 3D, improper ($\epsilon=0.5$) null point.
Regardless of the eccentricity, $\epsilon$, of the null point studied, we find that:
\begin{itemize}
\item {An initially pure, linear fast wave exhibits no linear or nonlinear coupling to the Alfv\'en mode in the neighbourhood of a 3D null point.}
\item {Due to the absence of coupling, the propagation of the fast wave is entirely dictated by the Alfv\'en-speed profile which about a 3D null point leads to a refraction effect,  focusing all the wave energy at the null point itself.}
\item {The propagating fast wave generates and sustains an instantaneous and dependent nonlinear field-aligned disturbance.} 
\end{itemize}

One of the chief aims of the simulations  was to look for evidence of MHD mode coupling about fully 3D (i.e. improper) null points. Unlike the proper \textit{symmetric} 2.5D null, it  was not clear from analysis of the MHD equations as to whether wave modes about improper 3D nulls are coupled due to the asymmetric magnetic topology, nor whether these coupling terms are linear or nonlinear, \emph{i.e. it was not clear what the overall general effect of departing from azimuthal symmetry would have on the wave dynamics.} To address this 
we have generated a pure fast magnetoacoustic wave in the vicinity of 3D nulls and analysed its subsequent propagation. We find that this wave exhibits no mode coupling to the Alfv\'en mode at either a linear or nonlinear scale about both the proper ($\epsilon=1$) and improper null point ($\epsilon=0.5$). Further experiments, namely $\epsilon=0.25$ and $\epsilon=0.75$ (values $\epsilon>1$ have analogues in the range $0<\epsilon<1$, and $\epsilon=0$ recovers the 2D null point), retain these results. Thus, we find that lack of mode conversion is a general feature of all potential null points, \emph{i.e. magnetic fieldline eccentricity does not facilitate fast to Alfv\'en mode conversion.}

We additionally find that, in both numerical experiments, the fast-mode pulse generates and sustains a field-aligned disturbance in $|\mathbf{v}_{\mathbf{B}_0}|$.   Our analysis shows that this field-aligned disturbance is not self-sustaining: it is an instantaneous {\emph{daughter}} disturbance resulting from the {\emph{progenitor}} fast wave. It is evident that the disturbance in $|\mathbf{v}_{\mathbf{B}_0}|$  does not act as a mode-conversion mechanism in the cases considered here, since no signal is ever generated in $|\mathbf{v}_\mathbf{A}|$,  and the disturbance has no feedback effect upon the main wave.

Due to the absence of mode coupling, the key features of fast wave behaviour in the neighbourhood of the $\epsilon=0.5$, improper null points are fundamentally the same as the behaviour in 2D, $\beta=0$ null point studies (e.g. McLaughlin \& Hood \cite{MH2004}), despite the inhomogeneous, fully 3D magnetic field. The difference between propagating fast waves about different potential null points is due \textit{only} to the differing Alfv\'en-speed profiles.
Thus, the propagation of the fast wave is found to be entirely dictated by such profiles, causing the refraction effect which, over time, focuses all of the wave energy at the null point. This is confirmed by our numerical simulations and supports the conclusions drawn from a 3D WKB approach in McLaughlin et al. (\cite{MFH2008}). Note, in the numerical simulations at large time, the pulse is so close to the null point that the resolution will eventually become inadequate. 
Nonetheless, in our simulations, over a finite, resolvable  time the wave energy accumulates in a small spatial region around the null point, resulting in an exponential steepening of current-gradients, hence resistivity will eventually become non-negligible, resulting in ohmic heating of the local coronal plasma via resistive dissipation (a conclusion directly carried-over from 2D studies, see McLaughlin et al. \cite{james2011a}). Thus, \emph{we conclude that 3D null points are locations for preferential heating by passing fast magnetoacoustic waves.}

Wave motions and null points are ubiquitous in the corona and so these fast wave-null interactions  are likely inevitable. The large body of theory, extended by the results presented here, indicates that this will result in localised heating events. Thus, the key question is \emph{does this make a signifigant contribution to coronal heating?} To answer this, clear observational evidence for MHD waves around coronal nulls is needed, a calculation of resultant heating, and a survey of the prevalence of such events. Such observations would require the detection of coronal nulls using co-temporal high spatial/temporal imaging and magnetograms to study oscillations in the vicinity of coronal null points. It is possible that the \emph{Atmospheric Imaging Assembly} and \emph{Helioseismic and Magnetic Imager} aboard SDO may permit such a study. 

We cannot, however, assume the  results presented here hold in the opposite case, i.e. a similar scenario for an initially pure Alfv\'en wave, follows suit. In fact, results of  Galsgaard et al. (2003) report this may not be the case.     Such an investigation will be the subject of a future paper.

Finally,  we highlight that sufficiently close to the null point, as magnetic induction drops off, there will be a region where the magnitudes of the sound speed and Alfv\'en speed become comparable: identified by Bogdan et al. (\cite{bogdan}) as the \emph{magnetic canopy}, \textit{viz.} the $\beta=1$ layer. This was investigated for a 2D null point by McLaughlin \& Hood (\cite{MH2006b}) and has the effect of introducing the \textit{slow magnetoacoustic wave} into the system, which is coupled to the fast wave (but both are still decoupled from the Alfv\'en wave). Thus the cold, $\beta=0$ plasma assumption does not completely capture the physics of MHD waves about real null points, which \emph{must} contain a $\beta=1$ layer near the null point. Thus, to understand wave dynamics about null points in the corona, it is necessary to extend the model presented here and see how these results are modified when the $\beta=0$ assumption is removed.


\begin{acknowledgements}

The authors acknowledge IDL support provided by STFC, and thank the referee for helpful comments. 
JOT acknowledges travel support provided by the RAS and the IMA, and a Ph.D. scholarship provided by Northumbria University. 
The computational work for this paper was carried out on the joint STFC and SFC (SRIF) funded cluster at the University of St Andrews (Scotland, UK).

\end{acknowledgements}


\begin{appendix}

\section{Analytical confirmation of decoupling at the proper null point.}\label{section:A}
In our simulations  we find no evidence of any mode coupling from the fast to the Alfv\'en mode.
In the case of the proper null point, as it assumes azimuthal symmetry, we can confirm this analytically by taking the $\hat{\vec{\theta}}$-component of equation (\ref{equation_9}) considered in standard cylindrical polars for $\epsilon=1$:
\begin{eqnarray}
\displaystyle \frac{\partial^2 v_\theta}{\partial t^2} &=& \left[-1 + r \frac{\partial}{\partial r} +r^{2}\frac{\partial^2}{\partial r^2}+ 4z \frac{\partial}{\partial z} +4z^{2} \frac{\partial^2}{\partial z^2} - 4rz\frac{\partial^2}{\partial r \partial z}\right] v_\theta \nonumber \\
& & - 2z\frac{\partial}{\partial z}\left[\frac{\partial}{\partial z}\left(v_\theta b_z - v_z b_\theta \right) - \frac{\partial}{\partial r}\left(v_r b_\theta -v_\theta b_r\right)\right] \nonumber \\
& & + \frac{\partial}{\partial r}\left[ r\frac{\partial}{\partial r }\left(v_\theta b_r -v_r b_\theta \right) + r\frac{\partial}{\partial z }\left(v_\theta b_z -v_z b_\theta \right) \right] \nonumber \\
& & + \left[ b_r \left(\frac{1}{r} -\frac{\partial}{\partial r} - r \frac{\partial ^2}{\partial r^2}+\frac{2z}{r}\frac{\partial}{\partial z} + 2z \frac{\partial ^2}{\partial r \partial z}\right)   \right. \nonumber \\
& & + \left. b_z \left(3\frac{\partial}{\partial z} +2z \frac{\partial ^2}{\partial z^2} -r \frac{\partial^2}{\partial r \partial z}\right)\right] v_\theta \nonumber \\
& &  + \left(\frac{2z}{r}v_r +2z \frac{\partial v_r}{\partial r} +2v_z +r\frac{\partial v_z}{\partial r}\right)\frac{\partial b_\theta}{\partial z}  \nonumber \\
& & + \left.\left(2v_r +2z\frac{\partial b_r}{\partial z} +r\frac{\partial v_z}{\partial z}\right)\frac{\partial}{\partial r}\left(rb_\theta\right) \right.\nonumber \\
& & + \left(\mathbf{\nabla}\cdot\mathbf{v}-\mathbf{v}\cdot\mathbf{\nabla}\right)\left(1+r\frac{\partial}{\partial r}-2z\frac{\partial}{\partial z}\right)b_\theta \nonumber \\
& & +\rho_1 \left[-1 + r \frac{\partial}{\partial r} +r^{2}\frac{\partial^2}{\partial r^2}+ 4z \frac{\partial}{\partial z} +4z^{2} \frac{\partial^2}{\partial z^2} - 4rz\frac{\partial^2}{\partial r \partial z}\right] v_\theta \nonumber \\
& & + \left[\left(\mathbf{\nabla}\times\mathbf{b}\right)\times\mathbf{B}_{0}  \cdot\mathbf{\nabla}\right]v_\theta
\label{eq:xdef} 
 \end{eqnarray}
 
Here, the linear terms $\mathcal{O}(\alpha)$ are shown on the first line, and nonlinear terms of $\mathcal{O}(\alpha^2)$ follow.  For the proper null $\hat{\vec{\theta}} \propto   \hat{\mathbf{A}}$,  and so we consider $\hat{\vec{\theta}}$ which acts as a proxy for $\hat{\mathbf{A}}$. For an initial disturbance that is purely a fast mode disturbance, we take $b_\theta$ and $v_\theta$ as initially zero, yielding  ${\partial^2 v_\theta}  / {\partial t^2} =0$ for all time. Hence, equation (\ref{eq:xdef}) states that if $b_\theta=v_\theta=0$ initially then $v_\theta=0$, and hence  ${\mathbf{v}_{\hat{\mathbf{A}}} } = {\bf{0}}$ for all time.  This provides analytical confirmation that \emph{about the proper null there is no fast mode to Alfv\'en mode conversion at either a linear or nonlinear level.}

\end{appendix}

\end{document}